\documentclass[12pt,preprint,epsfig]{aastex}

\shorttitle{Local Group distances and publication bias. II. M31 and beyond}
\shortauthors{Richard de Grijs and Giuseppe Bono}

\begin{document}

\title{Clustering of Local Group distances: publication bias or
  correlated measurements? II. M31 and beyond}

\author{
Richard de Grijs\altaffilmark{1,2} and
Giuseppe Bono\altaffilmark{3,4}
}

\altaffiltext{1} {Kavli Institute for Astronomy and Astrophysics,
  Peking University, Yi He Yuan Lu 5, Hai Dian District, Beijing
  100871, China}
\altaffiltext{2} {Department of Astronomy, Peking University, Yi He
  Yuan Lu 5, Hai Dian District, Beijing 100871, China}
\altaffiltext{3} {Dipartimento di Fisica, Universit\`a di Roma Tor
  Vergata, via Della Ricerca Scientifica 1, 00133, Roma, Italy}
\altaffiltext{4} {INAF, Rome Astronomical Observatory, via Frascati
  33, 00040, Monte Porzio Catone, Italy}

\begin{abstract}
The accuracy of extragalactic distance measurements ultimately depends
on robust, high-precision determinations of the distances to the
galaxies in the local volume. Following our detailed study addressing
possible publication bias in the published distance determinations to
the Large Magellanic Cloud (LMC), here we extend our distance range of
interest to include published distance moduli to M31 and M33, as well
as to a number of their well-known dwarf galaxy companions. We aim at
reaching consensus on the best, most homogeneous, and internally most
consistent set of Local Group distance moduli to adopt for future,
more general use based on the largest set of distance determinations
to individual Local Group galaxies available to date. Based on a
careful, statistically weighted combination of the main stellar
population tracers (Cepheids, RR Lyrae variables, and the magnitude of
the tip of the red-giant branch), we derive a recommended distance
modulus to M31 of $(m-M)_0^{\rm M31} = 24.46 \pm 0.10$ mag---adopting
as our calibration an LMC distance modulus of $(m-M)_0^{\rm LMC} =
18.50$ mag---and a fully internally consistent set of benchmark
distances to key galaxies in the local volume, enabling us to
establish a robust and unbiased, near-field extragalactic distance
ladder.
\end{abstract}


\section{Introduction}
\label{intro.sec}

The accuracy of extragalactic distance measurements ultimately depends
on robust, high-precision determinations of the distances to the
galaxies in the Local Group. This is, of course, the basis of the
concept of the astronomical ``distance ladder'' (for an up-to-date,
modern version of the distance ladder, see de Grijs 2013). In
particular, distance measurements to the Large Magellanic Cloud (LMC)
have played an important role in constraining the value of the Hubble
constant, $H_0$. The {\sl Hubble Space Telescope (HST)} Key Project
(HSTKP) on the Extragalactic Distance Scale (Freedman et al. 2001)
estimated $H_0 = 72 \pm 3$ (statistical) $\pm 7$ (systematic) km
s$^{-1}$ Mpc$^{-1}$. Most notably, the $\sim 10$\% systematic
uncertainty affecting their determination of $H_0$ was said to be
predominantly driven by the remaining systematic uncertainties in the
assumed distance to the LMC prevalent at that time (Freedman et
al. 2001; Schaefer 2008; Pietrzy\'nski et al. 2013).

Yet, the accuracy of LMC distance determinations has been dogged by
persistent claims of ``publication bias'' (e.g., Schaefer 2008, 2013;
Rubele et al. 2012; Walker 2012). Therefore, in de Grijs et al. (2014;
henceforth Paper I) we re-analyzed the full body of LMC distance
measurements published between 1990 and 2013. We concluded that strong
publication bias is unlikely to have been the main driver underlying
the clustering of many published LMC distance moduli. However, we
found that many of the published values were based on highly
non-independent tracer samples and analysis methods. In turn, this has
led to significant correlations among the body of LMC distances
published since 1990. The earlier conclusion that the tight clustering
of published values and the reduction in the spread observed in recent
years may have been due to publication bias was, in essence, based on
inappropriate application of the Kolmogorov--Smirnov (KS) test to a
data set that ultimately did not meet the requirements for such tests:
KS tests are only applicable to samples that consist of independent
and identically distributed values. In the context of LMC distance
measurements, both constraints are violated. These violations
originate, in particular, from progress in data analysis, enlarged
tracer samples, and improvements in both our theoretical understanding
and in the model implementation of theoretical stellar evolution
scenarios, which all drive down the uncertainties in the derived
parameters (including distance moduli) and lead to more consistent
results.

We now extend our distance range of interest to include published
distance moduli to M31, a few of its companion galaxies, and a few
other well-known Local Group members in order to assess whether or not
these measurements may be affected by publication bias or correlations
among the methods employed to obtain them (cf. Dalcanton et
al. 2012). More importantly, however, we aim at reaching consensus on
the best, most homogeneous, and internally most consistent set of
Local Group distance moduli to adopt for future, more general
use. Combined with the results of Paper I, we aim at determining
whether the distance scale in the Local Group as a whole may need
further revision. The present series of papers is based on the largest
set of distance determinations to Local Group galaxies available to
date, which we have assessed carefully in the context of an analysis
of population-specific properties and biases never done to the same
extent. This paper is organized as follows. In Section \ref{data.sec}
we outline our approach to compiling our database of distance
measurements. In Section \ref{m31dist.sec} we explore in detail the
trends, if any, in the distance measurements to M31. We pay particular
attention to the calibration relations used to arrive at the M31
distances based on variable star tracers: see Section
\ref{calibration.sec}. We subsequently focus our analysis on a number
of M31 companion galaxies, as well as a few other well-known Local
Group members (Section \ref{m31group.sec}). Finally, in Section
\ref{conclusions.sec} we summarize and place our results in the
broader context. We conclude with our recommendations as to which
distances to Local Group galaxies constitute a homogeneous, internally
consistent set.

\section{Distance Measurements to the M31 Group}
\label{data.sec}

\subsection{M31 distance determinations, 1918--2013}

To compile our M31 distance database, we relied entirely on the
NASA/Astrophysics Data System (ADS) compilation of the published
literature, in the absence of suitable comprehensive databases we
might use as our basis (but see Vilardell et al. 2006). We checked all
of the nearly 13,000 articles published as of late January 2014 and
which contained references to M31 in the NASA/ADS compilation for
references to newly determined distances to the galaxy. Our
comprehensive search resulted in a total of 168 distance measurements
to M31 or components associated with the galaxy's main body. We did
not include determinations of the distances to the large population of
accompanying dwarf galaxies (but see Section
\ref{groupmembers.sec}). Only very few ($\sim$5) of the more recent
determinations include separate analyses of the statistical and
systematic uncertainties in the published measurements. Our final
database, sorted as function of both publication date and distance
tracer, is available from
http://astro-expat.info/Data/pubbias.html. Its structure is similar
to that used for our LMC distances database presented in Paper I.

Among the 117 newly determined distance moduli published since 1990,
which we take as the period of interest for our main analysis, only
three appeared in non-refereed publications. For the same reasons as
justified in Paper I, we opted to retain these measurement so as to
avail ourselves of a complete publication record. Three articles (Rich
et al. 2005; Mackey et al. 2006; Perina et al. 2009) reported distance
measurements to individual globular clusters (GCs) associated with
M31. We calculated the average values for each of these three data
sets (containing 19, 4, and 11 GCs each, respectively), which we will
use for further analysis in the remainder of this paper. Specifically,
we obtained $(m-M)_0^{\rm M31} = 24.49 \pm 0.15$ mag, $24.40 \pm 0.16$
mag, and $24.42 \pm 0.21$ mag for the mean distances based on the
individual GC measurements reported in Rich et al. (2005), Mackey et
al. (2006), and Perina et al. (2009), respectively. The uncertainties
quoted above reflect the spread among the individual GC distances
(i.e., the depth of the M31 GC system), as well as the typical
photometric uncertainty, which we added in quadrature.

\subsection{Published distances to selected additional Local Group galaxies}
\label{groupmembers.sec}

Since we aim at establishing a robust set of benchmark distances
within the local volume, we selected a number of additional Local
Group members of different types. In addition to the third largest
spiral galaxy in the Local Group, M33, we selected M31's close,
late-type companion M32 (classified as a compact elliptical-type
galaxy), the dwarf spheroidal (dSph)/dwarf elliptical (dE) galaxy pair
NGC 147/NGC 185, as well as the dSph/dE galaxy NGC 205 (M110), the
dwarf irregular object IC 10, and the irregular galaxy IC 1613. All of
these galaxies host a variety of tracers that can be used for accurate
distance determinations---each affected by its own systematic
uncertainties---and cross-calibration among the different tracers. The
inclusion of IC 1613 is particularly interesting from the perspective
of metallicity differences: the galaxy's Cepheids, for instance, are
characterized by a significantly different metallicity compared with
their Galactic counterparts, $\Delta \mbox{[Fe/H]} \simeq 1$ dex (for
a discussion, see e.g. Majaess et al. 2009). The full database of
newly reported distance measurements to each of our sample galaxies
can be accessed at http://astro-expat.info/Data/pubbias.html.

At the time of the completion of our online database, in late January
2014, the NASA/ ADS contained 6008 articles that referred to M33, as
well as 2256, 576, 758, 1299, 998, and 1057 articles containing
references to M32, NGC 147, NGC 185, NGC 205, IC 10, and IC 1613,
respectively. Our exhaustive exploration of these $\sim$12,000
publications led to inclusion of a total of 131 newly reported
distances to M33 since records began in 1926, as well as 38 to M32
(since 1944), and 37, 54, 43, 46, and 145 distance estimates to,
respectively, NGC 147, NGC 185, NGC 205, IC 10, and IC 1613. The most
commonly available distance tracers include Cepheids (M33, NGC 205, IC
10, and IC 1613) and RR Lyrae (M32, M33, NGC 147, NGC 185, NGC 205,
and IC 1613) variable stars, as well as features associated with
bright giant stars, such as the level of the tip of the red-giant
branch (TRGB; M32, M33, NGC 147, NGC 185, NGC 205, and IC 1613) or the
red clump (RC; M32, M33, and IC 1613), and the technique of surface
brightness fluctuations (SBFs; M32, NGC 147, NGC 185, and NGC 205; for
IC 10, an equivalent approach was taken by Yahill et al. 1977). M33 is
the only object among this additional sample of Local Group galaxies
for which we have access to two types of independent, {\it geometric}
distance determinations, based on water masers (Greenhill et al. 1993;
Argon et al. 1998, 2004; Brunthaler et al. 2005) and on a single
O-type eclipsing binary (EB) system (Bonanos et al. 2006; Bonanos
2007, 2008).

\section{Trends in Distance Determinations to M31?}
\label{m31dist.sec}

\begin{figure}
\begin{center}
\includegraphics[width=\columnwidth]{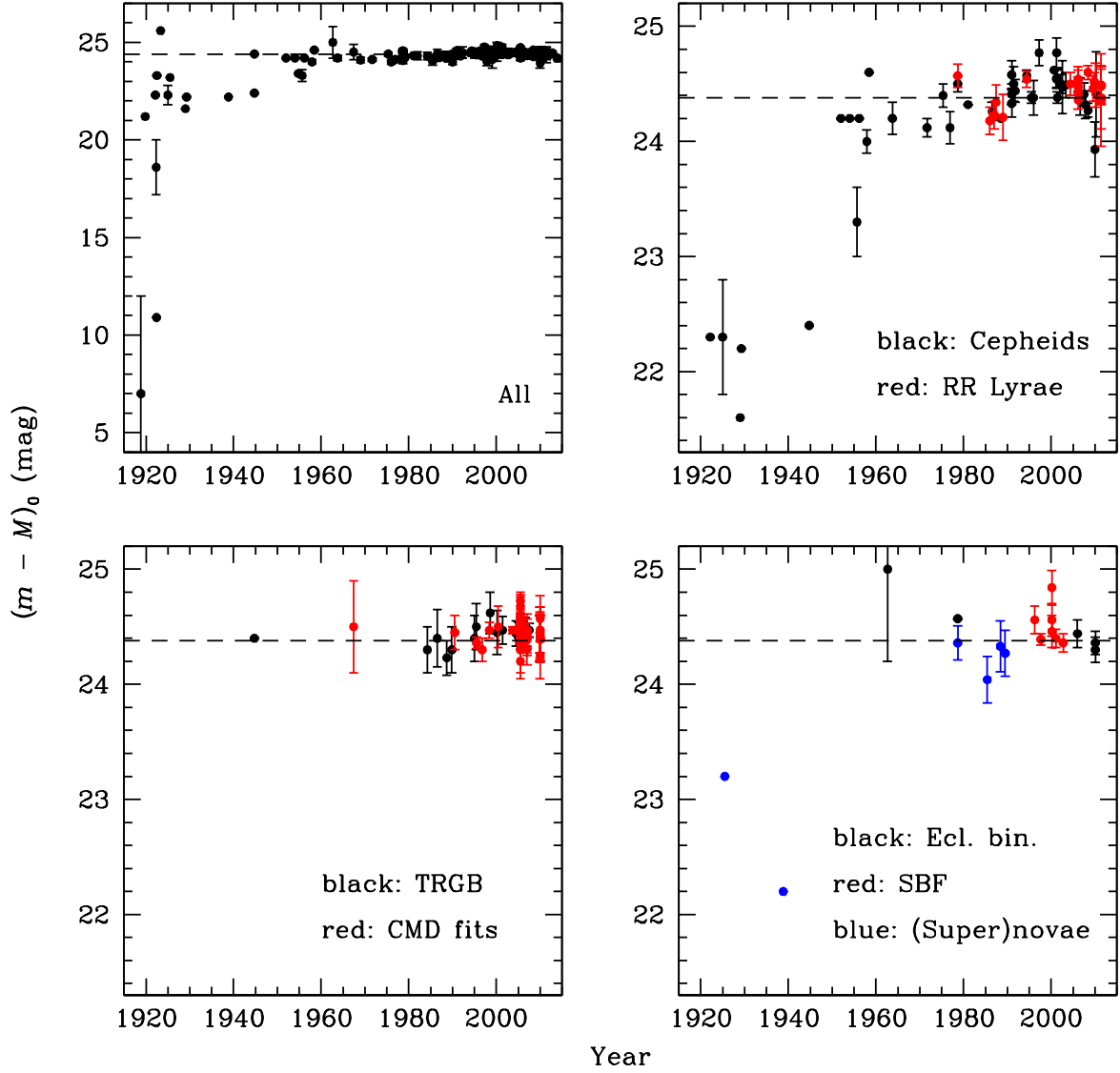}
\caption{Published extinction-corrected M31 distance moduli since
  records began as a function of publication date (year+month), where
  possible centered on the galaxy's center. The horizontal dashed
  lines indicate the ``canonical'' distance modulus of $(m-M)_0^{\rm
    M31} = 24.38$ mag (Freedman et al. 2001). TRGB: tip of the
  red-giant branch; CMD: color--magnitude diagram; Ecl. bin.:
  eclipsing binary systems; SBF: surface brightness fluctuations.}
\label{m31dist.fig}
\end{center}
\end{figure}

Figure \ref{m31dist.fig} shows the distribution of published distance
moduli, corrected for extinction by their respective authors, since
the first bold attempts by van Maanen (1918; with references to
earlier work), Lundmark (1919), and Luplau-Janssen \& Haarh (1922) to
measure a trigonometric parallax to the galaxy. Following subsequent
attempts to use galaxy dynamics to derive a distance (Jeans 1922;
\"Opik 1922), Hubble (1922, 1925a,b, 1929a,b) and Lundmark (1923,
1925; and references therein) were the first to use individual objects
in M31 as tracers of the system's distance as a whole. They used
classical novae and Cepheid variable stars, respectively, which were
easily accessible with telescopes that were available at the time
because of these objects' intrinsically high luminosities. It is,
therefore, not a surprise that to date the largest number of newly
reported distance measurements to M31 based on individual tracers (42)
are based on Cepheid calibrations.

A casual inspection of Figure \ref{m31dist.fig} reveals that the
distance to M31 has been known to within $\sim 15$\%, given the
sometimes sizeable uncertainties, since the 1960s. The average
distance modulus has been slowly increasing until the mid-1980s, when
it leveled off near a value of $(m-M)_0^{\rm M31} \simeq 24.4$ mag,
with a typical uncertainty of $\lesssim 0.15$--0.2 mag. In this paper,
we are particularly interested in exploring any more recent trends in
the average distance modulus, specifically during the ``modern''
period from 1990 until the present time. (The precise choice of
starting date of our modern period is not important, provided that we
have access to a sufficiently long time span that would allow us to
discern any statistical trends.) Therefore, we will not discuss the
M31 distance tracers based on nova- or supernova-related light-curve
features, since all of the latter were published prior to 1990. In the
remainder of this and in the next section, we will examine whether the
individual distance moduli published during this period may have been
based on either correlated data or model approaches, or perhaps been
subject to publication bias.

\begin{figure}
\begin{center}
\includegraphics[width=0.8\columnwidth]{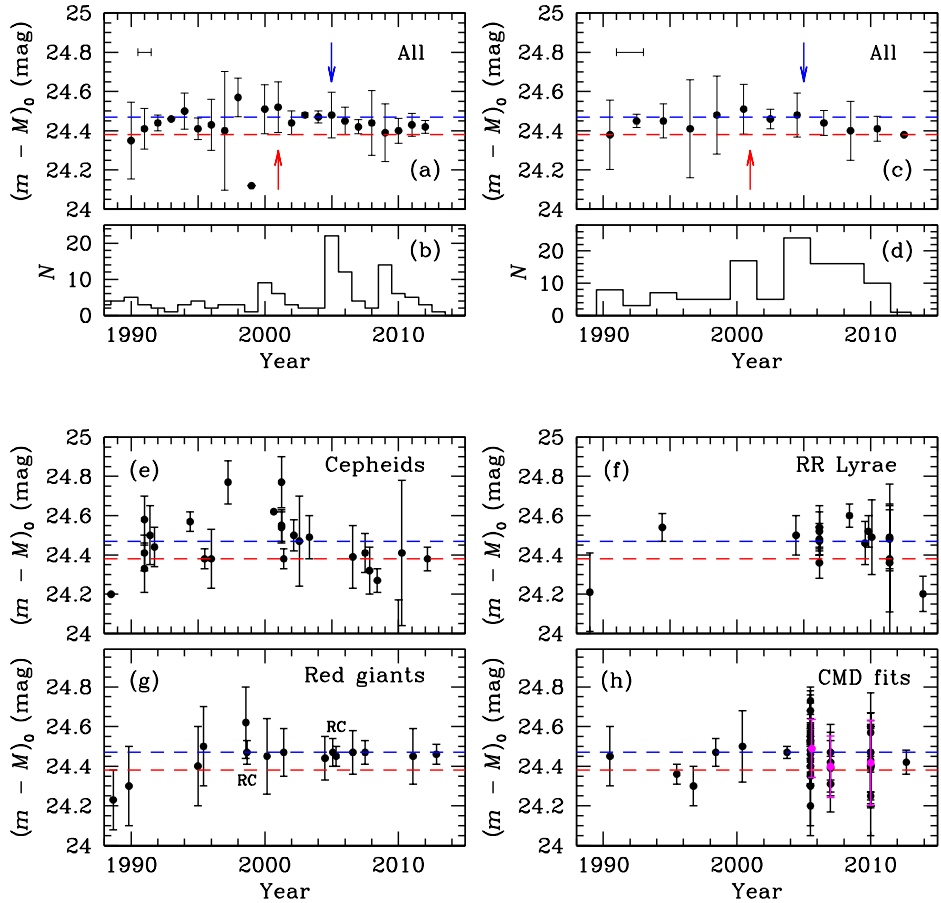}
\caption{Published M31 distance moduli since 1990. (a, b) and (c, d)
  Annual and biennial average distance moduli, respectively, based on
  the ensemble of distance tracers of all types, as well as the
  numbers of distance values considered. The red and blue dashed lines
  indicate the distance moduli published by Freedman et al. (2001; red
  arrow) and McConnachie et al. (2005; blue arrow), respectively. The
  horizontal ``error bars'' reflect the periods over which the
  individual data points have been averaged in the respective
  panels. (e)--(h) Individual distance measurements for the main
  tracer types. Panel (e) includes all Cepheid-based distances,
  including those based on classical Cepheids and the Type II
  Cepheid-based distance modulus of $(m-M)_0 = 23.93 \pm 0.24$ mag
  obtained by Majaess et al. (2009). The red giant-based distances in
  panel (g) are predominantly based on TRGB measurements, except for
  two values based on red clump (``RC'') observations. The magenta
  data points with their associated error bars in panel (h) indicate
  the galaxy-wide averages of the GC-based distance
  moduli---individually shown in this panel---of Rich et al. (2005),
  Mackey et al. (2006), and Perina et al. (2009), i.e., $(m-M)_0^{\rm
    M31} = 24.49 \pm 0.15$ mag, $24.40 \pm 0.16$ mag, and $24.42 \pm
  0.21$ mag, respectively, where the error bars include the spread
  among the individual data points as well as the typical photometric
  uncertainties.}
\label{m31dist2.fig}
\end{center}
\end{figure}

We follow a similar approach as in Paper I. In Figure
\ref{m31dist2.fig} we expose our time-restricted data set to further
scrutiny. The top panels in this figure show the annual and biennial
averages (as well as the number of data points considered) of all
distance measurements pertaining to the full period from 1990 until
2013. We specifically highlight the levels of the distance moduli
determined by Freedman et al. (2001), i.e., the ``canonical'' distance
modulus of $(m-M)_0^{\rm M31} = 24.38$ mag, and that of McConnachie et
al. (2005), $(m-M)_0^{\rm M31} = 24.47$ mag. We will use these
determinations as our benchmarks to assess the occurrence, if any, of
publication bias.

The arrows indicate the publication dates of our two benchmark
distance moduli, where the colors correspond to the relevant
horizontal dashed lines. At first sight, it does not appear that
following the publication of either of our benchmark distance moduli
the average levels converged to the respective values. Figures
\ref{m31dist2.fig}e through \ref{m31dist2.fig}h include the individual
distance measurements for four types of common distance tracers, i.e.,
Cepheid and RR Lyrae variable stars, red giants, and color--magnitude
diagram (CMD) fits. The individual distance moduli, sorted by tracer,
are available at http://astro-expat.info/Data/pubbias.html. Except
for the red giant-based distances, we do not discern any clear trends
among the individual measurements. As regards the red giants, although
their average level does not vary significantly as a function of
publication year, the associated uncertainties shrink quite
significantly from 1990 until the present time. The red giant-based
distances in Figure \ref{m31dist2.fig}g are predominantly based on
measurements of the tip of the TRGB magnitude, except for two values
based on RC observations. The observed reduction in the distance
moduli based on this distance tracer is reminiscent of the situation
for the LMC distance moduli based on RC data (Paper I). Of the 10
articles using the TRGB magnitude as their distance tracers,
six\footnote{Specifically, Morris et al. (1994), Couture et
  al. (1995), Salaris \& Cassisi (1998), Durrell et al. (2001),
  McConnachie et al. (2005), and Conn et al. (2012).}  are based on
newly obtained, independent observations. The remaining four articles
(Ferrarese et al. 2000; Sakai et al. 2004; Saha et al. 2006; Rizzi et
al. 2007) use a subset of observations discussed in the six
independent articles; all of these articles base at least part of
their analyses on the published data of Mould \& Kristian
(1986). Their calibration relations are largely independently
determined based on either empirical or theoretical (stellar
evolution) considerations.

We thus conclude that there is no compelling reason to assume a
significant contribution from publication bias to the bulk of
present-day M31 distance moduli. We suggest that the observed
clustering of the TRGB-based data points is likely related to our
improved understanding of the details of stellar evolution. This is
corroborated by the notion that the TRGB magnitudes used as
calibration benchmarks span a very narrow range in absolute $I$-band
magnitude. The main uncertainties associated with the use of the TRGB
as a distance indicator are therefore related to the way in which the
TRGB magnitude is detected. However, given that modern approaches are
based on very large numbers of stars, edge-detection techniques run
into few difficulties in this regard. This is exemplified by the small
scatter in the resulting distance estimates to M31 shown in Table
\ref{stats.tab}. In the next section, we will examine the distance
moduli resulting from the use of period--luminosity relations (PLRs)
of variable stars, as well as calibration issues specific to M31 and
the converging trends of the average distance moduli over time.

\section{Calibration}
\label{calibration.sec}

To assess whether the post-1990 Cepheid- and RR Lyrae-based distance
determinations to M31 have been largely independent or instead been
subject to significant correlations among data sets, zero-point
calibrations, and methods employed, we carefully examined the origins
of the individual distance determinations.

We considered 20 publications yielding M31 distance moduli based on
Cepheid light curves, as well as eight articles based on RR Lyrae
variables (for details, see our online database). Close examination of
these 28 papers revealed that more than two thirds (14 Cepheid- and
six RR Lyrae-based distance determinations) of recent M31 distance
estimates rely on the LMC's distance as their calibration
benchmark. One article, Majaess et al. (2009), reports a distance
modulus based on analysis of Type II Cepheid light curves; this
distance estimate is discrepant at the 1.5--2$\sigma$ level with
respect to all other modern Cepheid-based M31 distance
measurements. This is likely owing to the criterion these authors
adopted to distinguish between classical and Type II Cepheids. As they
show in their figures 1 and 2, in M31 the transition between the two
different types is rather smooth, thus rendering a clear distinction
between both Cepheid types uncertain. Because of this issue and the
intrinsically different nature of the objects considered by Majaess et
al. (2009), we will not include this result in the ensuing
discussion. (Note that although Type II Cepheids originate from the
same old stellar population as RR Lyrae stars, a comparison of this
latter data point with the mean value defined by the RR Lyrae distance
moduli leads to a similarly discrepant result.)

Although 12 of the 19 articles based on classical Cepheids use
$(m-M)_0^{\rm LMC} = 18.50$ mag as their calibration benchmark, the
LMC distance moduli adopted by the various authors range from
$(m-M)_0^{\rm LMC} = 18.42$ to 18.54 mag. This implies that the
Cepheid- and RR Lyrae-based distance estimates shown in Figure
\ref{m31dist.fig}a, c, e, and f are not all based on the same distance
scale. We thus proceeded to adjust the distance moduli of all Cepheid-
and RR Lyrae-based determinations to a common LMC benchmark of
$(m-M)_0^{\rm LMC} = 18.50$ mag. For the seven articles whose M31
distance estimates were not based on application of a relative
M31--LMC distance modulus,\footnote{{\it Cepheids}: Feast \& Catchpole
  (1997), Mochejska et al. (2000), Paturel et al. (2002a,b), and Riess
  et al. (2012); {\it RR Lyrae}: Sarajedini et al. (2009) and
  Fiorentino et al. (2010).} we checked the LMC distance moduli their
resulting distances would be equivalent to (many authors state these
values, in fact, while in a few cases this amounted to a simple
conversion). Note that the discrepant Type II Cepheid distance modulus
of $(m-M)_0 = 23.93 \pm 0.24$ mag cannot be made more consistent with
the distance moduli based on classical Cepheids by adjusting the
adopted LMC distance, $(m-M)_0^{\rm LMC} = 18.45$ mag, to our
benchmark value.

\begin{figure}
\begin{center}
\includegraphics[width=\columnwidth]{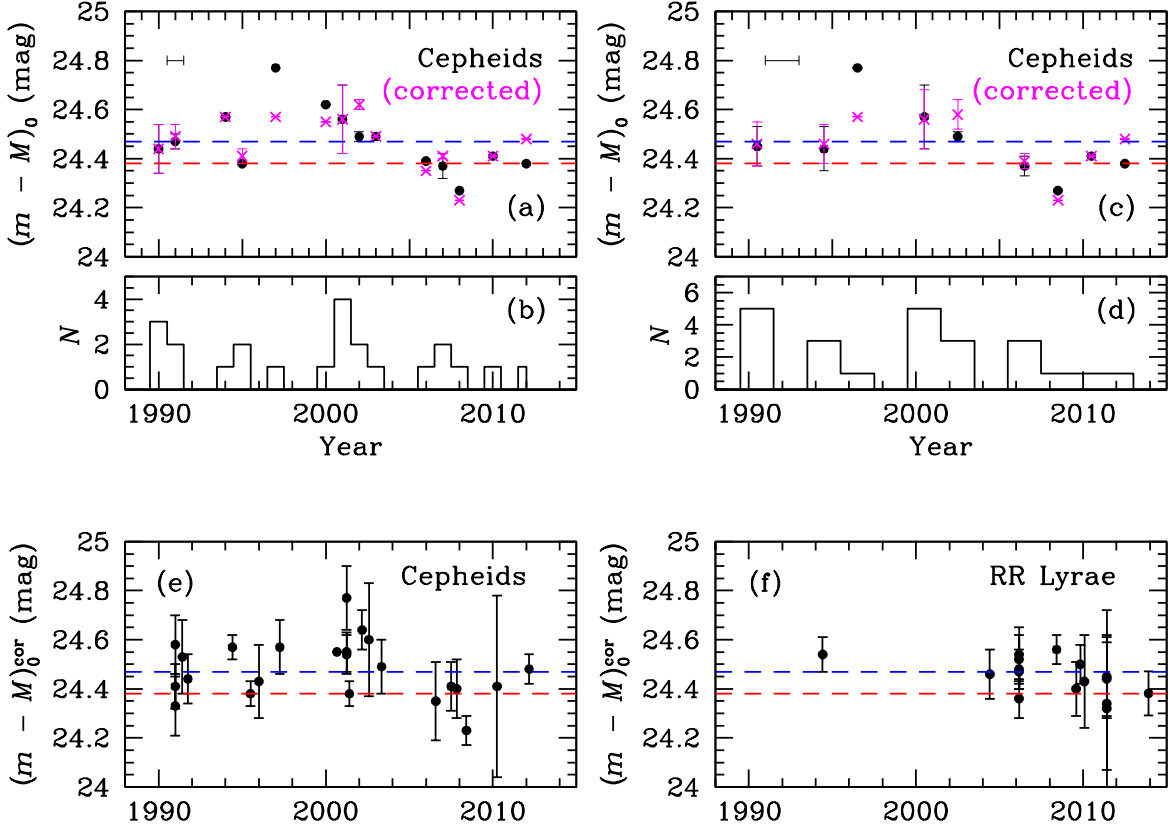}
\vspace{-2.5cm}
\caption{As Figure \ref{m31dist2.fig}, but for the variable star-based
  distance moduli to M31 only. The magenta data points in panels (a)
  and (c) have been corrected with respect to the original
  measurements (black points) to reflect a calibration based on the
  canonical LMC distance modulus of $(m-M)_0^{\rm LMC} = 18.50$
  mag. The individual Cepheid- and RR Lyrae-based distance moduli in
  panels (e) and (f) also reflect this adjustment; panel (e) includes
  only classical Cepheid-based distance moduli.}
\label{m31dist3.fig}
\end{center}
\end{figure}

Figure \ref{m31dist3.fig} is equivalent to Figure \ref{m31dist3.fig},
except that we have now ``corrected'' the relevant distance moduli to
a common LMC benchmark modulus of 18.50 mag. Figures
\ref{m31dist3.fig}a and \ref{m31dist3.fig}c show the size of this
effect in the context of the annual and biennial averages. Once again,
we do not see any significant trend(s) for either the Cepheid- or the
RR Lyrae-based distance determinations, although we note that there
may be some hints of publication bias in the results, as reflected by
the clustering---above or below the benchmark levels---of data points
published close in time (for Cepheid-based distances, cf. the periods
between 2000 and 2004, and from 2006 to 2009). In Table
\ref{stats.tab}, we compare the average M31 distance moduli for the
comparison period from early 1990 until May 2001 and for two periods
following the benchmark publications of Freedman et al. (2001) and
McConnachie et al. (2005), for both the original and the ``corrected''
values. We do not include an average based on RR Lyrae variables for
the comparison period prior to the publication of Freedman et
al. (2001), since only a single RR Lyrae-based distance estimate was
published during that period (Gould 1994), i.e., $(m-M)_0^{\rm M31} =
24.54 \pm 0.07$ mag for $(m-M)_0^{\rm LMC} = 18.50$ mag. In Section
\ref{conclusions.sec} we will discuss the broader implications of the
values listed in Table \ref{stats.tab}.

\begin{table}
\caption{Statistical properties of the body of M31 distance
  measurements prior to and following publication of the two benchmark
  values used in this paper. ``Corrected'' values refer to distance
  moduli calibrated to the ``canonical'' LMC distance modulus,
  $(m-M)_0^{\rm LMC} = 18.50$ mag. Means and population standard
  deviations are given in units of magnitudes.}
\label{stats.tab}
\begin{center}
\begin{tabular}{@{}llccccccccc@{}}
\hline \hline
          &          & \multicolumn{8}{c}{Period} \\
\cline{3-10}
          &          & \multicolumn{2}{c}{00/1990--05/2001$^a$} && \multicolumn{2}{c}{06/2001--12/2013} && \multicolumn{2}{c}{02/2005--12/2013} \\
\cline{3-4}\cline{6-7}\cline{9-10}
          &          & Orig.  & Corr.   && Orig.   & Corr.     && Orig.   & Corr. \\
\hline
          & Mean     & 24.454 &         && 24.435  &           && 24.432  \\
All       & $\sigma$ &  0.164 &         &&  0.110  &           &&  0.115  \\
          & $N$      & 42     &         &&  76     &           &&  68     \\
\hline
Classical & Mean     & 24.516 & 24.502  && 24.404  & 24.446    && 24.363  & 24.380 \\
Cepheids  & $\sigma$ &  0.140 &  0.115  &&  0.072  &  0.118    &&  0.052  &  0.077 \\
          & $N$      & 14     & 14      &&  9      &  9        &&  6      &  6     \\
\hline
          & Mean     &        &         && 24.457  & 24.443    && 24.454  & 24.442 \\
RR Lyrae  & $\sigma$ &        &         &&  0.097  &  0.072    &&  0.100  &  0.075 \\
          & $N$      &        &         && 15      & 15        && 14      & 14     \\
\hline
          & Mean     & 24.517 & 24.505  && 24.437  & 24.444    && 24.427  & 24.424 \\
All variables & $\sigma$ & 0.135 & 0.111&&  0.092  &  0.093    &&  0.097  &  0.081 \\
          & $N$      & 15     & 15      && 24      & 24        && 20      & 20     \\
\hline
          & Mean     & 24.488 &         && 24.462  &           && 24.467 \\
TRGB      & $\sigma$ &  0.082 &         &&  0.012  &           &&  0.006 \\
          & $N$      &  5     &         &&  5      &           && 3      \\
\hline \hline
\end{tabular}
\end{center}
\flushleft
$^a$including Freedman et al. (2001).
\end{table}

\section{M31 Group Members}
\label{m31group.sec}

We will now discuss the distance estimates pertaining to M33 and IC
1613 separately, because of the fairly large numbers of distance
estimates available, followed by a general exploration of the
distances reported for the remaining sample galaxies.

\subsection{Distance determinations to M33 and IC 1613}

\begin{figure}
\begin{center}
\includegraphics[width=\columnwidth]{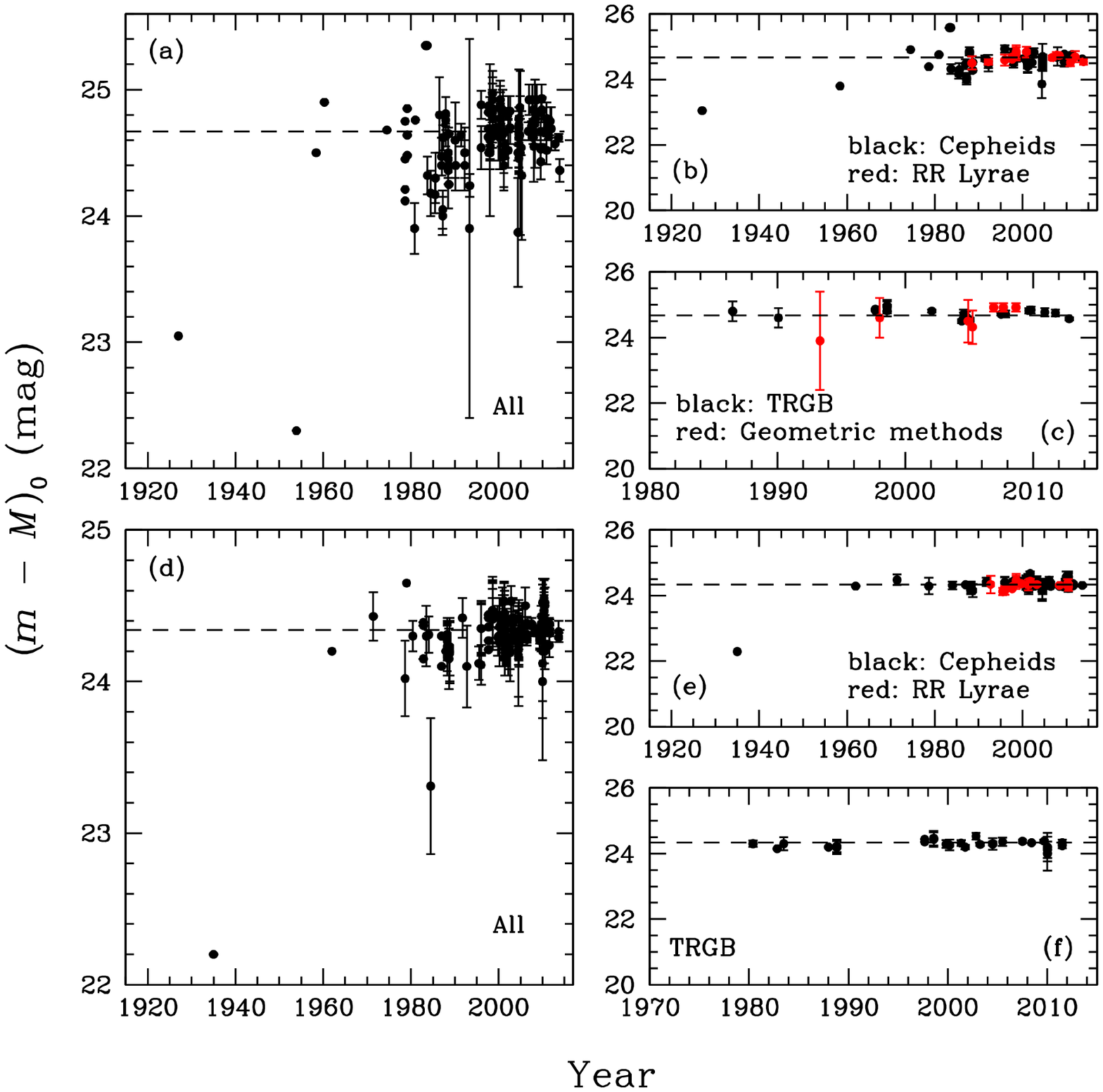}
\caption{As Figure \ref{m31dist.fig}, but for (a)--(c) M33 and
  (d)--(f) IC 1613. The distance moduli based on Cepheids and RR Lyrae
  variables---panels (b) and (e)---have been corrected to a distance
  scale defined by a canonical LMC distance modulus of $(m-M)_0^{\rm
    LMC} = 18.50$ mag. The horizontal dashed lines indicate our
  proposed ``best'' distance moduli of $(m-M)_0 = 24.67$ mag (for M33)
  and $(m-M)_0 = 24.34$ mag (for IC 1613), respectively (see the
  text).}
\label{m33dist.fig}
\end{center}
\end{figure}

Figure \ref{m33dist.fig} is equivalent to Figure \ref{m31dist.fig},
except that it relates to distance estimates to M33 and IC 1613. Among
the different distance tracers available for M33, those based on
Cepheids ($N = 56$, of which three are Type II Cepheids), RR Lyrae
variables ($N = 14$), and the level of the TRGB ($N = 20$) are most
numerous. Similarly, the distance determinations to IC 1613 based on
Cepheids ($N = 84$; two Type II measurements), RR Lyrae variables ($N
= 14$), and the TRGB magnitude ($N = 30$; including a small number of
other giant-based distance determinations published in the 1980s)
dominate the total tally. We will thus focus our discussion on these
tracers.

As for M31, we first updated the calibration of the distance estimates
based on Cepheids and RR Lyrae variables (as well as those based on
long-period variables, for M33) relative to an LMC distance modulus of
$(m-M)_0^{\rm LMC} = 18.50$ mag. Figures \ref{m33dist.fig}a and
\ref{m33dist.fig}d display all distance determinations we collected,
without application of any corrections; panels b, c, e, and f show the
tracer-specific data sets. We refer the reader to our discussion in
Section \ref{other.sec} for a detailed analysis of the trends and
properties shown in Figure \ref{m33dist.fig}.

\subsection{Comments on our remaining sample galaxies}

\begin{figure}
\begin{center}
\includegraphics[width=\columnwidth]{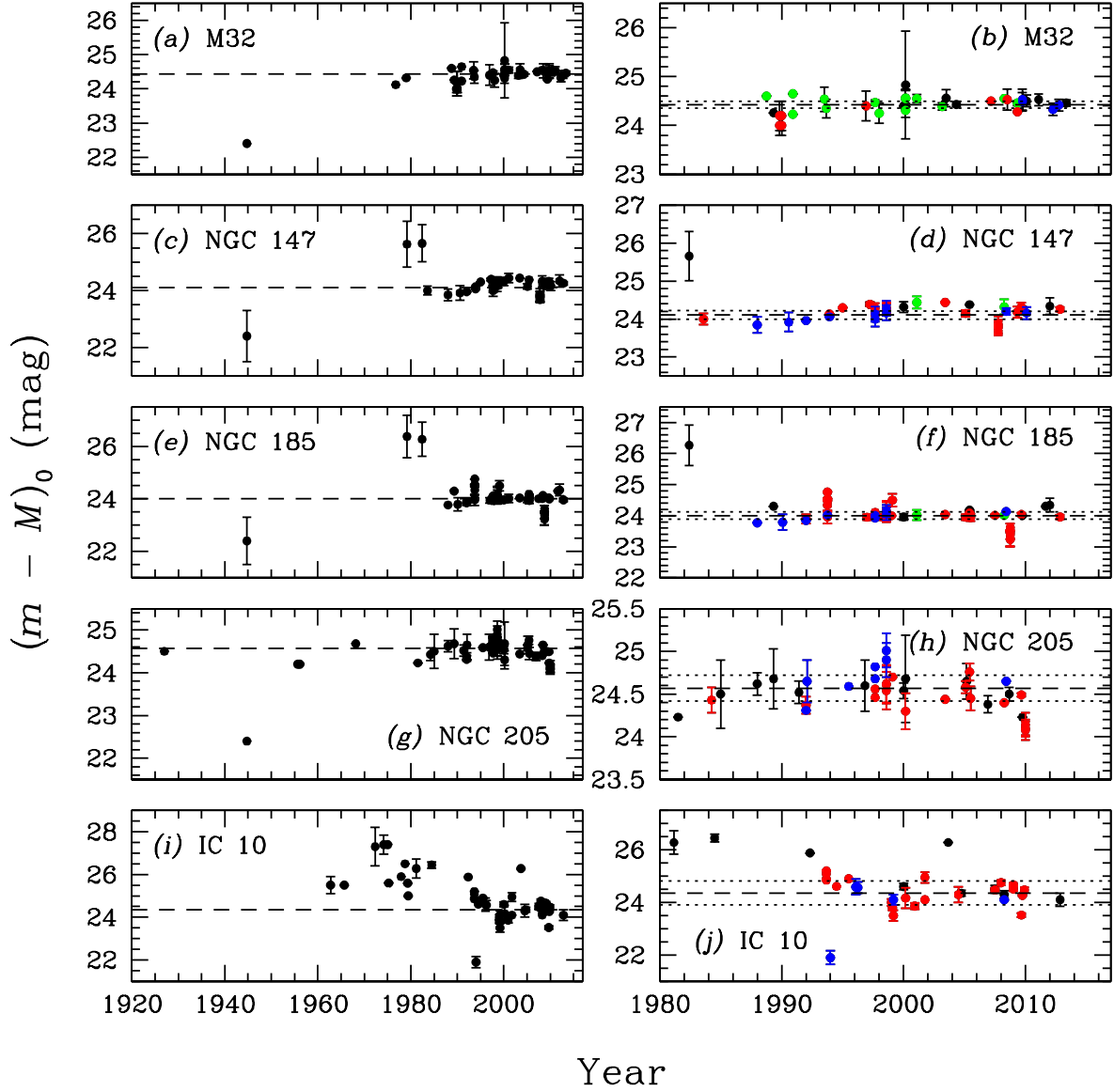}
\caption{As Figure \ref{m31dist.fig}, but for (a) and (b) M32, (c) and
  (d) NGC 147, (e) and (f) NGC 185, (g) and (h) NGC 205, and (i) and
  (j) IC 10. (left) Full chronological trends in distance
  determinations for each of our sample galaxies. (right) Zooms of the
  more recent timeframe since 1980. Red: TRGB-based estimates. Blue:
  RR Lyrae-based distance determinations (Cepheid-based distances for
  IC 10). Green: SBF distances. The horizontal dotted lines represent
  the $1\sigma$ uncertainties on the weighted mean distance moduli
  (for IC 10, they reflect the uncertainties in the TRGB-based
  distances).}
\label{dwarfs.fig}
\end{center}
\end{figure}

The other five galaxies in our small sample of Local Group member
systems only have between 37 and 54 distance determinations each, in
all cases dominated by a small number of tracers. Therefore, we will
discuss them jointly in this section. Figure \ref{dwarfs.fig} shows
the full chronological trends in distance determinations for each of
our galaxies (left-hand column), as well as zooms of the more recent
timeframe since 1980 (right-hand column). In the latter panels, we
have color-coded the distances resulting from the dominant tracer
populations: red data points originate from distance measurements
based on the TRGB method, blue points come from variable stars (RR
Lyrae variables; Cepheid-based distance estimates for IC 10), and
green points represent estimates based on the SBF method. The latter
was originally developed on the basis of (ground-based) observations
of M32 (Tonry \& Schneider 1988), although more recent estimates use
{\sl HST}-based imaging observations at red optical or near-infrared
(NIR) wavelengths (cf. Ferrarese et al. 2000). Except for the 13
SBF-based data points for M32, we also have two such data points for
both members of the NGC 147/NGC 185 galaxy pair (Tonry et al. 2001;
Tully et al. 2008). We will discuss the implications of the trends and
properties seen in Figure \ref{dwarfs.fig} in detail in Section
\ref{other.sec}.

\section{Verdict}
\label{conclusions.sec}

\subsection{A consensus M31 distance modulus}

In Section \ref{calibration.sec} we pointed out that one needs to
carefully correct the M31 distance determinations to define a common
distance scale. We discussed the nature of these corrections and
showed the results in Figure \ref{m31dist3.fig} and Table
\ref{stats.tab}. Although the scatter among the modern, post-2001
values in Table \ref{stats.tab} is non-negligible, the combination of
all 34 post-2001 values pertaining to the Cepheid and RR Lyrae
variables (corrected to a common LMC distance modulus of 18.50 mag),
as well as the TRGB distances, leads to a robust M31 distance modulus
of $(m-M)_0^{\rm M31} = 24.46 \pm 0.10$ mag. The post-2001 and
post-2005 averages for the Cepheid and RR Lyrae variable stars, as
well as those for our TRGB comparison sample, comfortably fall within
the mutual uncertainty ranges. At the distance of M31, the
line-of-sight location of the individual distance tracers with respect
to the galaxy's midplane is negligibly affected by depth issues,
except for members of the M31 GC system. This latter effect is
exemplified by the spread in distance moduli among the galaxy's GCs as
determined by Rich et al. (2005), Perina et al. (2009), and---to a
lesser extent (because of the smaller number of GCs included)---Mackey
et al. (2006): see Figure \ref{m31dist2.fig} (bottom right-hand
panel). We note in passing that Clementini et al.'s (2009) robust
distance measurement to the M31 GC B154---$(m-M)_0^{\rm B154} = 24.52
\pm 0.08$ mag---is formally consistent with the consensus distance
modulus derived here. This was, in fact, the first robust distance
derivation that was based on distance determinations to a large sample
of 89 RR Lyrae variables in a single M31 GC. The offset between both
measurements most likely reflects the GC's position at a distance that
is slightly greater than that to the galaxy's center. This is
supported by the spread in GC distances implied by the results of Rich
et al. (2005), Mackey et al. (2006), and Perina et al. (2009).

Salaris \& Cassisi (1998) reported a systematic difference between
their distance moduli---based on a theoretical calibration of the TRGB
magnitude using the models of Salaris \& Cassisi (1997)---and the
Cepheid-based empirical distance scale of Lee et al. (1993), which is
based on observations of Galactic GCs hosting RR Lyrae stars. They
found that the TRGB scale yields longer distances by 0.12 mag; the
offset does not seem to depend on metallicity. Salaris \& Cassisi
(1998) suggest that this systematic difference underscores the need
for a revision of the zero point of the Cepheid distance scale. On the
basis of the average distance moduli listed in Table \ref{stats.tab},
we find a systematic difference of the same order, 0.08--0.10 mag, for
the post-2005 average distance moduli (although we note that this may
be owing to small-number statistics). Mochejska et al. (2000) explored
the effects of blending of Cepheids with intrinsically luminous stars
in the disk of M31 using a combination of ground- and space-based
({\sl HST}) images (see also Vilardell et al. 2007). They concluded
that the Cepheid distance scale {\it pertaining to M31} requires a 9\%
upward adjustment to counteract crowding effects, which is
approximately twice the systematic difference found both by Salaris \&
Cassisi (1998) and in this paper (Table \ref{stats.tab}).

It is instructive to compare our recommended M31 distance modulus of
$(m-M)_0^{\rm M31} = 24.46 \pm 0.10$ mag with the most ``direct''
distance moduli obtained to date. In the latter case, we refer to the
geometric distances based on observations of EB systems. The most
recent distance estimates to M31 based on EBs yield distance moduli
ranging from $(m-M)_0^{\rm M31} = 24.44 \pm 0.12$ mag (Ribas et
al. 2005) to $(m-M)_0^{\rm M31} = 24.36 \pm 0.10$ (Vilardell et
al. 2010a,b). We note that the most recent value, which is based on a
combination of the individual distance determinations to two EB
systems, is somewhat smaller than our recommended value (although
still within the mutual $1\sigma$ uncertainties). This is again
reminiscent of the situation we encountered for the LMC in Paper I,
where the EB distances based on hot, early-type were systematically
smaller than those resulting from cool, late-type (and longer-period)
EBs. We attributed this to the more significant systematic
uncertainties affecting the hotter systems (cf. Pietrzy\'nski et
al. 2013).

Finally, we comment on the recent determination of the distance
modulus to M31 by Riess et al. (2012), $(m-M)_0^{\rm M31} = 24.38 \pm
0.06 \mbox{ (statistical)} \pm 0.03$ (systematic) mag. These authors
highlight their high-precision result, with an unprecedented total
uncertainty of 3\%. However, in the context of the discussion in
Section \ref{calibration.sec}, we point out that they adopted a
distance scale corresponding to an LMC distance modulus of
$(m-M)_0^{\rm LMC} = 18.40$ mag. While their precision may indeed be
unprecedented, to fit on the distance scale that has emerged in the
course of our work---and on which we report in both Paper I and
here---their corrected distance modulus is $(m-M)_0^{\rm M31} = 24.48
\pm 0.07$ mag (total uncertainty). This comfortably falls within the
$1\sigma$ uncertainties of our recommended value.

\subsection{Self-consistent distances to the M31 group}
\label{other.sec}

Distance determinations to M33 exhibit a large spread compared with
other satellites of the M31 group. The full range spans some 30\% with
respect to the mean and depends on the type of distance indicator. A
robust distance estimate was recently published by the Araucaria
Project (Gieren et al. 2013) based on the NIR PLR defined by two dozen
long-period classical Cepheids. They derived a true distance modulus
$(m-M)_0^{\rm M33} = 24.62 \pm 0.03$ (statistical) $\pm 0.06$
(systematic) mag and a reddening of $E(B-V) = 0.19 \pm 0.02$
mag. These same authors provide a detailed analysis of the
uncertainties that might affect these estimates, including their
dependence on metal abundance (Bono et al. 2010; Bresolin 2011),
crowding, and---in particular---reddening estimates. The latter seem
to represent a thorny problem, since recent estimates based on either
an O-type EB (Bonanos et al. 2006) or blue supergiant stars for which
individual spectroscopic determinations were obtained (U et al. 2009)
yield reddening estimates that differ by a factor of two. M33 is the
only ``dwarf'' spiral in the Local Group and there is no doubt that it
is a very interesting laboratory to constrain possible systematic
errors affecting both old (e.g., RR Lyrae variables) and young (e.g.,
Cepheids) solid distance indicators.

\begin{table}
\caption{Statistical properties of the key M33 and IC 1613 distance
  measurements. Distance moduli based on Cepheid and RR Lyrae variable
  stars have been calibrated to the ``canonical'' LMC distance
  modulus, $(m-M)_0^{\rm LMC} = 18.50$ mag. Means and population
  standard deviations are given in units of magnitudes.}
\label{stats_m33.tab}
\begin{center}
\begin{tabular}{@{}llccccc@{}}
\hline \hline
          &          & \multicolumn{2}{c}{M33} && \multicolumn{2}{c}{IC 1613}\\
\cline{3-4}\cline{6-7}
          &          & 00/1990 -- & 00/2000 -- && 00/1990 -- & 00/2000 -- \\
          &          & 12/1999    & 12/2013    && 12/1999    & 12/2013 \\
\hline
Classical & Mean     & 24.620 & 24.573 && 24.410 & 24.360 \\
Cepheids  & $\sigma$ &  0.157 &  0.188 &&  0.011 &  0.107 \\
          & $N$      & 10     & 25     &&  6     & 59     \\
\hline
          & Mean     & 24.680 & 24.677 && 24.291 & 24.341 \\
RR Lyrae  & $\sigma$ &  0.116 &  0.114 &&  0.121 &  0.062 \\
          & $N$      &  6     &  7     &&  8     &  6     \\
\hline
          & Mean     & 24.833 & 24.698 && 24.407 & 24.289 \\
TRGB      & $\sigma$ &  0.128 &  0.108 &&  0.068 &  0.118 \\
          & $N$      &  6     & 13     &&  6     & 17     \\
\hline
{\bf Weighted} & {\bf Mean} & {\bf 24.719} & {\bf 24.671} && {\bf 24.409} & {\bf 24.336} \\
           & {\bf $\sigma$} & {\bf  0.075} & {\bf  0.072} && {\bf  0.011} & {\bf  0.049} \\
\hline \hline
\end{tabular}
\end{center}
\end{table}

In Table \ref{stats_m33.tab}, we provide the statistical properties of
the distances to M33 and IC 1613 resulting from our key tracers, split
into decade-long time intervals. The exact period ranges adopted for
these intervals are not important; our aim here is to check whether
there may have been significant shifts in the ``best'' distance moduli
for both galaxies since the early 1990s. The ``best-fitting'' distance
moduli indicated by the horizontal dashed lines in Figure
\ref{m33dist.fig} are the weighted means from the most recent period
considered in Table \ref{stats_m33.tab}. The differences between the
weighted mean distance moduli for each of our three tracer populations
in the period since 2000 are smaller than the relevant $1\sigma$
uncertainties.\footnote{Note that the apparent ``trend'' for M33 of
  increasing distance modulus from Classical Cepheids to RR Lyrae
  variables and the TRGB is opposite that for IC 1613. This, combined,
  with the $\sim 1 \sigma$ variation among the mean distance moduli
  for all three tracer populations, indicates that these ``trends''
  are not physically real (because in many cases the calibration
  relations underlying the results were similar or the same) but
  simply reflect persistent statistical uncertainties.}

It is instructive to compare our recommended distance moduli for M33
and IC 1613---i.e., $(m-M)_0^{\rm M33} = 24.67 \pm 0.07$ mag and
$(m-M)_0^{\rm IC\,1613} = 24.34 \pm 0.05$ mag---with those implied by
alternative methods of distance determination that have not
contributed to these ``best'' values. For M33, we have access to a
number of ``direct,'' geometric methods ({\it H$_{\it 2}$O masers}:
Argon et al. 2004; Brunthaler et al. 2005; {\it O-type EB}: Bonanos et
al. 2006; Bonanos 2007, 2008; note that these latter values are not
independent). Figure \ref{m33dist.fig}c shows that our recommended
distance modulus for the galaxy is encompassed by the error bars of
all geometric methods. Similarly, a comparison of the post-2000
distance moduli and their published uncertainties, derived based on
fits to CMD features ({\it RC}: Kim et al. 2002; Orosz et al. 2007;
{\it CMD fits}: Barker et al. 2011), the Tully--Fisher relation (Tully
et al. 2008), and the high-luminosity cutoff to the planetary nebulae
luminosity function (PNLF; Magrini et al. 2000; Ciardullo et
al. 2004), shows good agreement within the statistical
uncertainties. On the other hand, two types of distance tracers lead
to systematic differences in their resulting best values, although
both pertain to only a single publication each: the long-period
variable calibration of Pierce et al. (2000) leads to a systematic
difference of order $2 \sigma$, while the novel flux-weighted
gravity--luminosity relation also leads to a systematically larger
distance estimate (cf. U et al. 2009) at the 2--$3 \sigma$ level.

For IC 1613 the three post-2000 distance estimates based on RC
magnitudes (Dolphin et al. 2001, 2003; Udalski et al. 2001) are fully
consistent with our recommended value. We also point out that
Scowcroft et al. (2013) derived a weighted average of 57 distance
estimates from the NASA Extragalactic Database (NED) that is
consistent with our result, i.e., $(m-M)_0^{\rm IC\,1613} = 24.33 \pm
0.07$ mag, although the NED data set is incomplete at the time of
writing and their approach does not allow control of
population-specific systematic uncertainties, as we have attempted
here.

\begin{table}
\caption{Statistical properties of the key distance measurements to
  the additional five Local Group dwarf galaxies covering the
  timeframe from 1990 until the end of 2013. Distance moduli based on
  Cepheid and RR Lyrae variable stars have been calibrated to the
  ``canonical'' LMC distance modulus, $(m-M)_0^{\rm LMC} = 18.50$
  mag. Means and population standard deviations are given in units of
  magnitudes.}
\label{dwarfs.tab}
\begin{center}
\begin{tabular}{@{}llccccc@{}}
\hline \hline
Galaxy  & Tracer     & Mean & $\sigma$ & $N$ \\
\hline
M32     & SBF        & 24.451 & 0.133 & 12 \\
        & TRGB       & 24.318 & 0.201 &  6 \\
        & RR Lyrae   & 24.443 & 0.088 &  4 \\
        & {\bf Mean} & 24.430 & 0.069 \\
\hline
NGC 147 & TRGB       & 24.155 & 0.223 & 17 \\
        & RR Lyrae   & 24.098 & 0.120 &  9 \\
        & {\bf Mean} & 24.111 & 0.106 \\
\hline
NGC 185 & TRGB       & 24.027 & 0.333 & 26 \\
        & RR Lyrae   & 23.993 & 0.128 &  8 \\
        & {\bf Mean} & 23.997 & 0.119 \\
\hline
NGC 205 & TRGB       & 24.447 & 0.200 & 18 \\
        & RR Lyrae   & 24.701 & 0.214 &  8 \\
        & {\bf Mean} & 24.565 & 0.146 \\
\hline
IC 10   & {\bf TRGB} & 24.355 & 0.451 & 20 \\
        & Cepheids   & 23.852 & 1.117 &  5 \\
\hline \hline
\end{tabular}
\end{center}
\end{table}

In Table \ref{dwarfs.tab} we offer statistical insights into the
robustness and internal consistency of the distance determinations to
each of remaining five dwarf galaxies for the key tracer
populations. We also provide the weighted-mean, recommended distance
moduli for further use, where we have taken into account the number of
data points contributing to each mean tracer-based distance estimate,
the intrinsic spreads in those determinations, and the individual
error bars pertaining to the published values. The variable star-based
distance moduli have been rescaled to reflect an LMC distance modulus
of $(m-M)_0^{\rm LMC} = 18.50$ mag. Again, we adopted the period since
1990 (and until the end of December 2013) as our ``modern'' timeframe
for further analysis. For IC 10, we adopted as ``best'' distance
modulus that based on the TRGB only because of the unusually large
scatter in the Cepheid-based distances published for this galaxy since
1990.

We have indicated the $1\sigma$ uncertainty levels on the weighted
mean distance moduli in the right-hand panels of Figure
\ref{dwarfs.fig} (using the mean TRGB as a proxy of the weighted mean
for IC 10). The uncertainties are small for M32, NGC 147, and NGC
185. For M32, the most obvious outliers are two low TRGB-based
distance estimates from Freedman (1990), who provided their best
estimates for the distance to M32 based on two different
identifications of the TRGB level (determined using the brightest and
second brightest discontinuities in the luminosity function of the red
giant branch), at $(m-M)_0^{\rm M32} = 24.0$ mag and 24.2 mag,
respectively. Based on our analysis of the metadata for M32, we
recommend a ``best'' distance modulus of $(m-M)_0^{\rm M32} = 24.43
\pm 0.07$ mag. This recommendation is also supported by the three
independent RC-based distance estimates published for the galaxy
(Worthey et al. 2004; Fiorentino et al. 2010; Monachesi et al. 2011),
yielding a mean of $\langle (m-M)_0^{\rm M32} \rangle = 24.49$ mag and
a $1\sigma$ uncertainty of 0.04 mag. The data point with the large
error bar in Figure \ref{dwarfs.fig}b is based on calibration of the
bright cutoff of the PNLF (Ferrarese et al. 2000), yielding
$(m-M)_0^{\rm M32, PNLF} = 24.83 \pm 1.10 \pm 0.13$ (systematic)
mag. We note that, because of the close association of M32 with and
its projection onto the disk of M31, until the dynamical modeling of
Byrd (1976), all previous articles that needed to adopt a distance to
M32 simply took the galaxy to be located at the distance of M31.

The published distance moduli for both NGC 147 and NGC 185 have
reached an approximately stable level in recent years. Notable
outliers occurred, however (NGC 147: Kang et al. 2009, only provided
as an abstract; NGC 185: Sohn et al. 2008). Both articles, published
by a subset of the same authors, provided TRGB-based estimates that
were systematically lower than the long-term average values. These
systematically smaller distance estimates may be owing to a
combination of the authors' use of the Yonsei--Yale isochrones for
their calibration instead of the more often used Padova isochrones and
their NIR $JHK$ calibration instead of the more customary $I$-band (or
equivalent) calibration. This difference gives, therefore, a useful
quantitative indication as regards the remaining systematic
uncertainties in (red) optical versus NIR TRGB calibration. For both
galaxies, post-1990 independent distance tracers---such as those based
on SBFs, horizontal-branch stars (Han et al. 1997; Butler \&
Mart\'{\i}nez-Delgado 2005), ($K$-band) long-period variables (Lorenz
2011; Lorenz et al. 2011, 2012), and kinematics-based methods
(Devereux et al. 2009)---lead to estimated distance moduli that are
fully consistent with our weighted means.

Figures \ref{dwarfs.fig}g--j (and in particular panels h and j) show
that the distance estimates to NGC 205 and IC 10 continue to be
subject to larger fluctuations among published values. Their distance
estimates appear to have converged to some extent in the recent past,
however. The clear, systematically low set of distance outliers for
NGC 205 were published by Jung et al. (2009), whose results deviate to
a similar extent from the bulk of the measurements and the long-term
average, $(m-M)_0^{\rm NGC\,205} = 24.57 \pm 0.15$ mag, as the
estimates of Kang et al. (2009) and Sohn et al. (2008) for NGC 147 and
NGC 185, respectively. The Jung et al. (2009) article shares a large
subset of the same authors and is based on the same approach as these
other two papers. In addition, where Devereux et al. (2009) found
distances in line with the long-term average for NGC 147 and NGC 185,
for NGC 205 their distance estimate is systematically lower, at
$(m-M)_0^{\rm NGC\,205} \sim 24.23$ mag, than the long-term
average. This may indicate lingering systematic effects caused by
peculiar motions due to the dominant presence of M31.

Among the M31 satellite galaxies, IC 10 plays a key role, since it is
a very actively star-forming galaxy. It is, in fact, considered the
only analog to the so-called ``post-starburst'' dwarf galaxies in the
Local Group (Gil de Paz et al. 2003). However, it is located at low
Galactic latitude, and its photometry is severely affected by
foreground extinction. This partially explains the broad range in
distance estimates associated with IC 10. These range from
$(m-M)_0^{\rm IC\,10} = 23.5 \pm 0.2$ mag ($D \simeq 0.5$ Mpc), based
on application of the TRGB method (Sakai et al. 1999), to well beyond
the Local Group using the PNLF (Jacoby \& Lesser 1981), $(m-M)_0^{\rm
  IC\,10} = 26.28 \pm 0.45$ mag ($D \simeq 1.8$ Mpc). A similarly
significant scatter is seen in relation to the galaxy's reddening
estimates, which range from $E(B-V)=0.8$ mag (Richer et al. 2001) to
$E(B-V)=1.2$ mag (Sakai et al. 1999). More recently, Sanna et
al. (2008)---using a new calibration of the TRGB method and very
accurate {\sl HST}/Advanced Camera for Surveys photometry---derived
new estimates of both the distance, $(m-M)_0^{\rm IC\,10} = 24.56 \pm
0.08$ (statistical) $\pm 0.08$ (systematic) mag, and the galaxy's
reddening, $E(B-V)= 0.78 \pm 0.06$ mag. This, in turn, supports
additional extant evidence that IC 10 is a likely member of the M31
subgroup.

Indeed, the set of distance estimates to IC 10 shows a larger number
of outliers with respect to the long-term average than the other M31
group members discussed above, although a consensus distance seems to
have been reached in the period since 2000. In this most recent
period, one TRGB-based distance (Kim et al. 2009) is clearly
discrepant (i.e., low) with respect to the mean. A second discrepant
(i.e., large) distance is, in fact, an {\it upper limit} resulting
from PNLF analysis (Magrini et al. 2003). More recent PNLF-based
determinations (Kniazev et al. 2008; Gon\c{c}alves et al. 2012) are
fully consistent with the weighted mean for this galaxy. Other notable
distance tracers in the post-2000 period, particularly those based on
carbon stars (Demers et al. 2004; Vacca et al. 2007), are also in line
with our expectations for the bulk of the data set.

\subsection{The bigger picture}

We are now in a good position to make a series of recommendations for
the use of robust distance measurements to a set of key Local Group
galaxies. We provide a summary of our basic recommendations, based on
both Paper I and the present work, in Table \ref{recommendations.tab}.

\begin{table}
\caption{Recommended distance moduli (as a function of increasing
  distance) to selected Local Group galaxies, comprising a robust
  local framework.}
\label{recommendations.tab}
\begin{center}
\begin{tabular}{@{}lclcc@{}}
\hline \hline
Galaxy   & $(m-M)_0^{\rm best}$ (mag) & Tracer(s) & $(m-M)_0^{\rm TRGB}$ (mag) & Ref.$^a$ \\
\hline
LMC$^b$  & $18.49 \pm 0.09$ & Cepheids, RR Lyrae, CMD  & 18.54--18.69     & I,$^*$ \\
NGC 185  & $24.00 \pm 0.12$ & TRGB, RR Lyrae           & $24.03 \pm 0.33$ & II \\
NGC 147  & $24.11 \pm 0.11$ & TRGB, RR Lyrae           & $24.16 \pm 0.22$ & II \\
IC 1613  & $24.34 \pm 0.05$ & Cepheids, RR Lyrae, TRGB & $24.29 \pm 0.12$ & II \\
IC 10    & $24.36 \pm 0.45$ & TRGB                     & $24.36 \pm 0.45$ & II \\
M32      & $24.43 \pm 0.07$ & SBF, TRGB, RR Lyrae      & $24.32 \pm 0.20$ & II \\
M31      & $24.45 \pm 0.10$ & Cepheids, RR Lyrae, TRGB & $24.47 \pm 0.01$ & II \\
NGC 205  & $24.56 \pm 0.15$ & TRGB, RR Lyrae           & $24.45 \pm 0.20$ & II \\
M33      & $24.67 \pm 0.07$ & Cepheids, RR Lyrae, TRGB & $24.70 \pm 0.11$ & II \\
NGC 4258 & $29.29 \pm 0.08$ & H$_2$O masers            & 29.24--29.44     & H99,$^*$ \\
\hline \hline
\end{tabular}
\end{center}
{\flushleft {\sc Notes:}\\ 
$^a$ I, II: Papers I, II; H99: Herrnstein et al. (1999; see text);
  $^*$ Various authors have determined TRGB-based distances to the LMC
  and NGC 4258. The ranges in distance moduli reflect the body of
  published work since 2000. LMC measurements based on the TRGB
  magnitude were published by Romaniello et al. (2000), Sakai et
  al. (2000), and Bellazzini et al. (2004); NGC 4258 measurements
  include Mouhcine et al. (2005), Rizzi et al. (2007), Mager et
  al. (2008), and Madore et al. (2009).\\
$^b$ All variable star-based distances reported in this table have
  been rescaled to the recommended LMC distance modulus determined in
  Paper I, $(m-M)_0 = 18.49 \pm 0.09$ mag. In practice, this has only
  affected (through a shift by $-0.01$ mag) the ``best'' distance
  moduli to M31 and NGC 205.  }
\end{table}

\begin{figure}
\begin{center}
\includegraphics[width=\columnwidth]{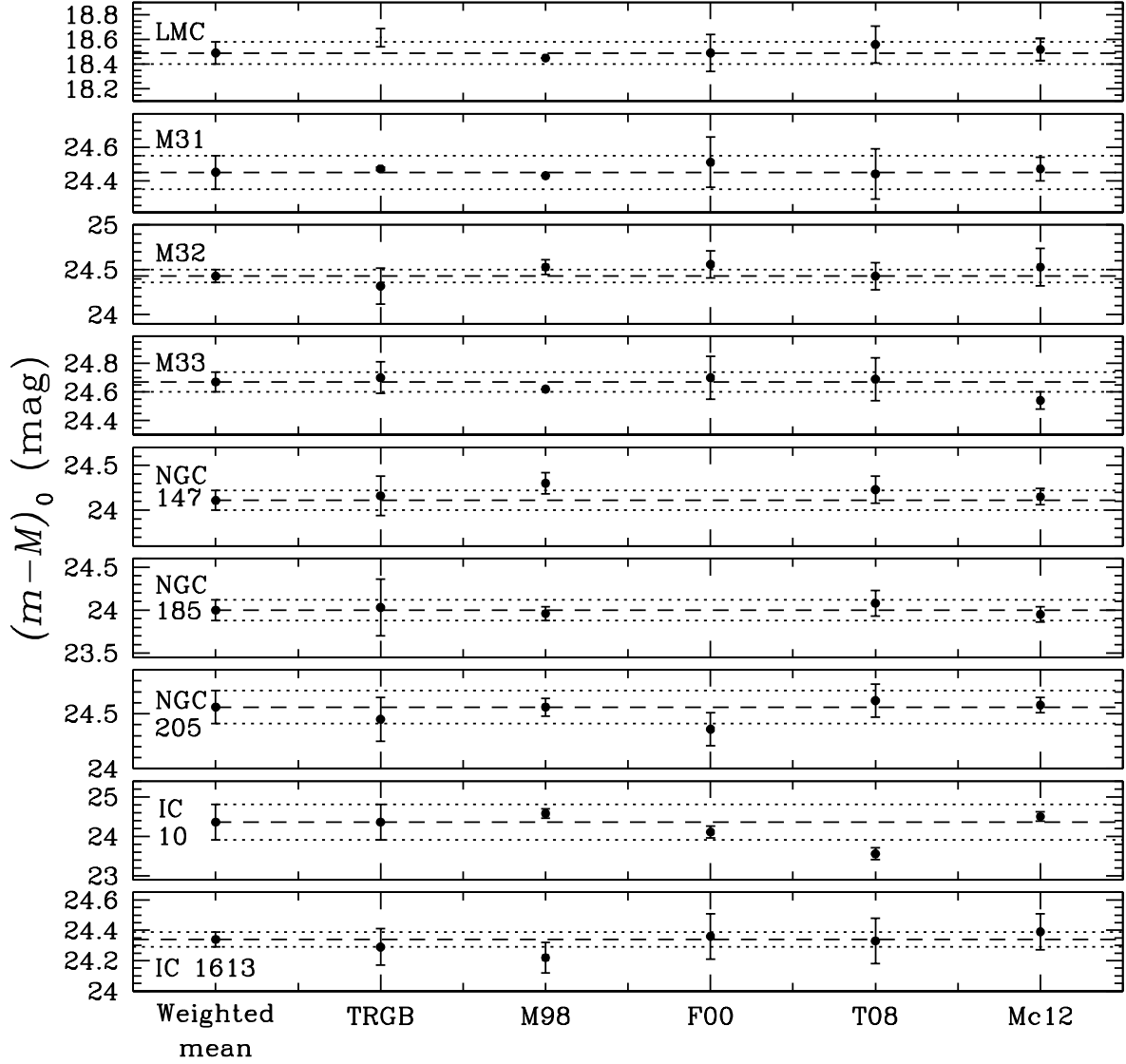}
\caption{Comparison of our set of benchmark distances to the sample of
  Local Group galaxies (indicated in the individual panels) discussed
  in this article with those from a number of recent distance
  compilations. ``Weighted mean'' and TRGB: this work, except where
  indicated in Table \ref{recommendations.tab}; M98: Mateo (1998);
  F00: Ferrarese et al. (2000); T08: Tammann et al. (2008); Mc12:
  McConnachie (2012).}
\label{litcf.fig}
\end{center}
\end{figure}

In Figure \ref{litcf.fig} we compare our set of benchmark distances to
our sample of Local Group galaxies with those from a number of recent
distance compilations. Because TRGB-based distances are the common
denominator among all of our sample galaxies, we show both the
weighted mean distance moduli derived in this paper (and in Paper I
for the LMC) and those based on the TRGB in the respective sample
galaxies. For comparison, we also show the mean levels and the
$1\sigma$ spreads in distance modulus implied for all galaxies. It is
clear that all comparison data sets exhibit significant scatter in the
relative distance moduli between many possible choices of pairs of
sample galaxies, even considering the published error
bars.\footnote{Note that Tammann et al. (2008) adopted a generic,
  generous uncertainty of 0.15 mag for all of their weighted mean
  distance moduli.} We remind the reader that our weighted means (as
well as the TRGB-based distances) are based on the largest data set of
distance measurements available to date, with particular emphasis on
converging trends in more recent years and a careful analysis of the
contributions from the different tracer populations. The data set that
exhibits the closest match to our set of benchmark distances is that
published by McConnachie (2012), although his distance to M33 is
significantly shorter than our recommended value. Through the body of
work presented in both Paper I and the present article, we aimed at
providing an updated, robust set of benchmark distances. Given our
adopted approach, the recommended values therefore supersede those
suggested in the context of older compilations, which were often based
on smaller metadata samples and earlier calibration attempts.

Going to near- and mid-IR wavelengths may enable us to reduce the
uncertainties in the distances to Local Group galaxies. At present,
2--3\% distance accuracy is already achievable to the LMC, and this
may be improved to $\sim 1$\% in the near future! For instance, the
Carnegie Hubble Program, using data from the warm {\sl Spitzer}
mission, derived $(m-M)_0^{\rm LMC} = 18.477 \pm 0.034$ mag (Freedman
et al. 2012). Meanwhile, Ripepi et al. (2012) used NIR {\sl VISTA}
observations to arrive at $(m-M)_0^{\rm LMC} = 18.46 \pm 0.03$ mag and
Inno et al. (2013) found $(m-M)_0^{\rm LMC} = 18.45 \pm 0.02$
(statistical) $\pm 0.10$ (systematic) mag based on optical/NIR PLR
analysis of a large sample of fundamental-mode LMC Cepheids. These
distances are comfortably close to and within the mutual uncertainties
of the direct, geometric distance determination based on eclipsing
binaries by Pietrzy\'nski et al. (2013), $(m-M)_0^{\rm LMC} = 18.493
\pm 0.008$ (statistical) $\pm 0.047$ (systematic) mag.

Water maser measurements, which were first applied to NGC 4258
(Herrnstein et al. 1999; see also Table \ref{recommendations.tab}),
have been extended to other nearby systems. Initial efforts to
determine the distance to M33 have thus far resulted in $D_{\rm M33} =
750 \pm 140 \pm 50$ kpc---$(m-M)_0^{\rm M33} = 24.38^{+0.49}_{-0.64}$
mag (total uncertainty)---where the first uncertainty in the linear
distance determination is related to uncertainties in the H{\sc i}
rotation model adopted for the galaxy, and the second uncertainty
comes from the proper motion measurements (cf. de Grijs 2013).

The technique of Very Long Baseline Interferometry is also
increasingly used to measure extragalactic proper motions. In turn,
this enables geometric distance determination out to some 100 Mpc,
including to the nearby galaxies NGC 4258, M33, UGC 3789, and NGC
6264. Combined with {\it a priori} information on a galaxy's
inclination with respect to our line of sight and its rotation curve,
based on radial velocity measurements, we can construct an accurate,
slightly warped ``tilted-ring'' model of the galaxy's dynamical
structure, usually assuming circular orbits (although this assumption
does not result in major systematic uncertainties). This, in turn,
allows correlation of the angular proper motion measurements with the
rotational velocity information obtained in linear units and, thus,
provides an independent distance measurement.

Simultaneously, the Megamaser Cosmology Project (e.g., Reid et
al. 2009, 2013; Braatz et al. 2010) aims at using extragalactic maser
sources to directly measure $H_0$ in the Hubble flow, which is clearly
a very challenging endeavor at distances in excess of 100 Mpc! Their
preliminary results look promising, however: using NGC 6264 ($D = 144
\pm 19$ Mpc) as a benchmark, they find $H_0 = 68 \pm 9$ km s$^{-1}$
Mpc$^{-1}$ (Kuo et al. 2013), which is indeed very close to the
current best determinations of $H_0$ based on a variety of independent
measures. This thus looks like a promising way forward to eventually
build a robust distance ladder out to the Hubble flow.

\begin{acknowledgements}
RdG is grateful for research support from the National Natural Science
Foundation of China through grants 11073001 and 11373010. RdG also
acknowledges research support from the Royal Netherlands Academy of
Arts and Sciences (KNAW) under its 2013 Visiting Professors
Programme. This work was partially supported by PRIN--INAF 2011 (PI
M. Marconi) and by PRIN--MIUR (2010LY5N2T, PI F. Matteucci). GB thanks
the Carnegie Observatories for support as a science visitor. This
research has made extensive use of NASA's Astrophysics Data System
Abstract Service.
\end{acknowledgements}

\end{document}